\documentclass[pra,aps,twocolumn,superscriptaddress,showpacs]{revtex4}
\usepackage{graphicx}
\usepackage{amsmath}

\newcommand{\eq}{\begin{equation}}
\newcommand{\fine}{\end{equation}}

\begin{document}

\title{
\vspace{2cm
\begin{flushright}
CERN-PH-TH/2015-035
\end{flushright}
\vspace{3cm}
\bf \LARGE Froissart Bound on Inelastic Cross Section \\ Without Unknown Constants   }
\vspace{.2cm}}
\author{Andr\'e Martin }
\email{martina@mail.cern.ch} \affiliation{Theoretical Physics Division,CERN, Geneva}

\author{S. M. Roy}
\email{smroy@hbcse.tifr.res.in} \affiliation{HBCSE,Tata Institute of Fundamental Research, Mumbai}

\begin{abstract}
Assuming  that axiomatic local field theory results hold for hadron scattering, Andr\'e Martin and S. M. Roy recently  
obtained absolute bounds on the D-wave below threshold for pion-pion scattering and thereby determined the 
scale of the logarithm in the Froissart bound on total cross sections in terms of pion mass only. Previously,  
Martin proved a rigorous upper bound on the inelastic cross-section $\sigma_{inel}$ which is one-fourth of the 
corresponding upper bound on $\sigma_{tot}$, and Wu, Martin,Roy and Singh improved the bound by adding the constraint of 
a given $\sigma_{tot}$ .

Here we use unitarity and analyticity to determine, without any high energy approximation, upper bounds on energy-averaged 
inelastic cross sections in terms of low energy data in the crossed channel. These are Froissart-type bounds without any unknown coefficient or  
unknown scale factors and can be tested experimentally. Alternatively, their asymptotic forms ,together with the Martin-Roy absolute 
bounds on pion-pion D-waves below threshold, yield absolute bounds on energy-averaged inelastic cross sections. E.g. 
for $\pi^0 \pi^0$ scattering , defining $\sigma_{inel}=\sigma_{tot} -\big (\sigma^{\pi^0 \pi^0 \rightarrow \pi^0 \pi^0} + 
\sigma^{\pi^0 \pi^0 \rightarrow \pi^+ \pi^-} \big )$ ,we show that for c.m. energy $\sqrt{s}\rightarrow \infty $,  
$\bar{\sigma}_{inel }(s,\infty)\equiv s\int _{s} ^{\infty } ds'\sigma_{inel }(s')/s'^2   \leq (\pi /4) (m_{\pi })^{-2} 
[\ln (s/s_1)+(1/2)\ln \ln (s/s_1) +1]^2$ where $1/s_1= 34\pi \sqrt{2\pi }\>m_{\pi }^{-2}  $ . This bound is  
asymptotically one-fourth of the corresponding Martin-Roy bound on the total cross section , and the scale factor 
$s_1$ is one-fourth of the scale factor in the total cross section bound. The average over the interval (s,2s) of 
the inelastic $\pi^0 \pi^0 $cross section has a bound of the same form with $1/s_1$ replaced by $1/s_2=2/s_1 $.
   
\end{abstract}


\maketitle

{\bf I. Introduction}. 
Recently we  \cite{Martin-Roy2014} have obtained bounds on energy averages of the total
cross-section without any unknown constants such as an overall constant factor or the scale factor in 
the logarithm. The purpose of the present work is to obtain analogous bounds on the energy averaged 
inelastic cross section without any unknown constants.The background and the basic postulates are again summarized 
below to make this work self-contained.

 The Froissart \cite{Froissart1961} bound on the total cross-section $\sigma_{tot} (s)$ for 
two particles at c.m. energy $\sqrt s$ ,
\begin{equation}
\sigma_{tot} (s) \leq_{s\rightarrow \infty} C \> [\ln (s/s_0)]^2 ,
\end{equation}
(where $C, s_0$ are unknown constants ) was initially proved assuming the Mandelstam representation .
This assumption might not be valid , for example , if there are rising Regge trajectories. Fortunately,  
 \cite{Martin1966} it was possible to prove this bound rigorously in the much more general frame work of 
Wightman's \cite{Wightman} axiomatic local field theory as applied to hadrons. Later, the needed analyticity 
properties , and polynomial boundedness at fixed momentum transfer squared $t$, were obtained by Epstein, Glaser and Martin 
\cite{Epstein} in the even more general framework of the theory of local observables of Haag, Kastler and Ruelle 
\cite{Haag}. It has nevertheless been questioned \cite{Azimov} if these properties apply to hadrons made of quarks and 
gluons. Zimmermann \cite{Zimmermann} has shown that local fields can be associated to composite particles.
We decide to believe that this proof applies to the present situation.
We postulate that the analyticity and polynomial boundedness derived from local field theory holds for hadrons .

In the proof of the Froissart bound in \cite{Martin1966} a crucial role is played by the use of unitarity to enlarge 
the Lehmann ellipse of analyticity \cite {Lehmann1958} for the absorptive part $A(s,t)$ to show that the 
right extremity $t_0$ of the enlarged ellipse in the $t$-plane stays non-zero and positive when $s\rightarrow \infty$.
For many processes, for example for 
 $\pi \pi ,KK, K\overline K,\pi K, \pi N, \pi \Lambda$ scattering it is known \cite{Bessis-Glaser1967} that
$t_0 = 4 m_{\pi}^2 $,,  $m_{\pi} $ being the pion-mass . (Except when especially necessary to show the 
dependence on pion-mass, we shall choose units $m_{\pi}=1$). Using unitarity and validity of 
dispersion relations with a finite number of subtractions for $-T< t\leq 0$ , Jin and Martin 
 \cite{Jin-Martin1964}) proved twice subtracted fixed-$t$ dispersion relations for $|t|< t_0 $. From this 
Lukaszuk and Martin \cite{Lukaszuk-Martin1967} fixed the constant $C$ in the Froissart bound to obtain,
\begin{equation}
 \sigma_{tot} (s) \leq _{s\rightarrow \infty}\sigma_{max} (s),
\end{equation}
 where,
 \begin{equation}
\sigma_{max} (s)  \equiv  4\pi/(t_0 - \epsilon ) \> [\ln (s/s_0)]^2 
\end{equation}
where  $\epsilon $ is an arbitrarily small positive constant. The Froissart-Martin bounds have inspired much work on high 
energy theorems (see e.g. \cite{Singh-Roy1970}, \cite{Roy1972}) and on models of high energy scattering \cite{Cheng-Wu1970}.
Further, Martin proved a  
bound on the total inelastic cross section $\sigma_ {inel} (s) $ at high energy \cite{Martin2009} which is 
one-fourth of the above bound $\sigma_{max} (s) $ on the total cross-section; and,
Wu, Martin, Roy and Singh \cite{WMRS} obtained a bound on $\sigma_ {inel} (s) $  which improves that bound 
if $\sigma_ {tot} (s)$ is known,
\begin{equation}
 \sigma_ {inel} (s) \leq_{s\rightarrow \infty} \sigma_ {tot} (s) \bigl ( 1- \sigma_{tot} (s) / \sigma_{max} (s) \bigr ).
\end{equation}
 {\bf The motivation for getting a bound on the inelastic cross section is the almost general belief \cite{Cheng-Wu1970} that 
 at high energies hadron total cross sections cannot exceed twice the inelastic cross section}. Hence, gaining a factor of 4 in the 
 inelastic cross section gains a factor of 2 in the total cross section.
 
One shortcoming of these bounds from the standpoint of rigour is that \cite{Yndurain} they are deduced by assuming  
 that the absorptive part $A(s,t), 0\leq t< t_0 $ is bounded by Const.$ s^2 /\ln (s/s_0) $
for $s\rightarrow \infty $, whereas axiomatic field theory results  
only guarantee that 
\begin{equation}
 C(t) \equiv \int _{s_{th }} ^{\infty} ds A(s,t) /s^3 < \infty,\>0\leq t< t_0, 
\end{equation}
where $s_{th }$ is the $s$-channel threshold. From the practical point of view a more 
serious shortcoming is that they involve the unknown scale factor $s_0$ in the argument of the 
logarithm and the unknown arbitrarily small but non-zero constant $\epsilon $.

In the case of the total cross section, both these shortcomings were removed recently \cite{Martin-Roy2014}. 
Bounds on energy averages of the total
 cross-section were obtained in which the scale $s_0 $ is determined in terms of 
$C(t)$. $C(t)$ can be bounded rigorously in terms of pion mass alone for $\pi^0 \pi^0$ scattering . Thus we 
obtained  the absolute bound\cite{Martin-Roy2014}, 
\begin{eqnarray}
\bar{\sigma}_{tot }(s,\infty)  &\leq& \pi (m_{\pi })^{-2} 
[\ln (s/s_0)+(1/2)\ln \ln (s/s_0) +1]^2\nonumber\\
&+&O(\ln \ln (s/s_0)), \> \> s_0^{-1}=17\pi \sqrt{\pi/2 }\> m_{\pi }^{-2},
\end{eqnarray}
where,
\begin{equation}
 \bar{\sigma}_{tot }(s,\infty)\equiv s\int _{s} ^{\infty } ds'\sigma_{tot }(s')/s'^2 .
\end{equation}
We also obtained somewhat improved bounds by using the additional 
phenomenological inputs for the D-wave scattering length \cite{Colangelo2000} for pion-pion scattering.
 
We prove here analogous bounds on energy averages of the inelastic cross section.
We choose the same normalizations as in Martin-Roy  \cite{Martin-Roy2014}. 
$F(s,t)$ denotes an $ab\rightarrow ab$ scattering amplitude  at c.m. energy $\sqrt{s}$ 
and momentum transfer squared $t$  normalized for non-identical partcles $a,b$ such that 
the differential cross-section $ d \sigma /d \Omega (s,t) $ is given by $\bigl |4 F(s,t)/\sqrt{s}  \bigr |^2 $,
with $t$ being given in terms of the c.m. momentum $k$ and the scattering angle $\theta$ by the 
relation,
\begin{equation}
 t= -2 k^2 (1- \cos \theta ); \>  z \equiv \cos \theta= 1+ t / (2 k^2).
\end{equation}
Then, for fixed $s$ larger than the physical $s-$channel threshold, $F(s;\cos \theta ) \equiv F(s,t)$ is 
analytic in the complex $\cos \theta $ -plane inside the Lehmann-Martin ellipse with foci -1 and +1 and 
semi-major axis $\cos \theta _0 = 1+ t_0 /(2 k^2) $.
Within the ellipse ,in particular, for $ |t| < t_0 , F(s,t)$ has 
the convergent partial wave expansion,
\begin{equation}
  F(s,t)=\frac{ \sqrt s }{4k} \sum_{l=0}^{\infty} (2l+1)P_l (z)a_l (s),
\end{equation}

with the unitarity constraint,
\begin{equation}
 Im a_l (s) \geq | a_l (s) |^2 ,\> s\geq s_{th} \>.
\end{equation}
Correspondingly, the optical theorem gives, for non-identical partcles $a,b$,
\begin{eqnarray}
 \sigma _{tot} (s) = \frac{4 \pi}{k}Im \big (4F(s,0)/\sqrt s \big )\nonumber\\
 =\frac{4 \pi}{k^2} \sum _{l=0}^{\infty} (2l+1) Im a_l(s)\>.
\end{eqnarray}
For identical particles ,e.g. for $\pi^0 \pi^0$ scattering, or for pion-pion scattering with Iso-spin $I$, 
 the partial waves $ a_l (s) \rightarrow 2 a_l^I (s) $ in the partial wave expansion,i.e.
\begin{equation}
 F^I(s,t)=\frac{ \sqrt s }{4k} \sum_{l=0}^{\infty} (2l+1) 2 a_l^I (s)P_l (z),
\end{equation}
\begin{equation}
 \sigma _{tot}^I (s) =  \frac{4 \pi}{k^2} \sum _{l=0}^{\infty} (2l+1) 2 Im a_l^I(s) \>,
\end{equation}
and we have the same formula for the differential cross-section in terms of $F(s,t)$, and the same 
form of the unitarity constraint,$ Im a_l^I (s) \geq | a_l^I (s) |^2 ,\> s\geq 4 \> $, as for non-identical 
particles. At threshold, $F^I (4,0)= a_0^I $, the S-wave scattering length for Iso-spin $I$. 

It will be seen that proofs of the bounds for inelastic cross sections are considerably more involved than those for total cross sections,
but the basic principles are the same. We give detailed derivations for the case of non-identical particles $a\neq b$, and also quote 
the identical particle results when needed.

{\bf II. Convexity Properties of Lower Bound on Absorptive Part in terms of Total Inelastic Cross Section}. 
We obtain a lower bound on the absorptive part of $F(s,t)$ (for $s\geq s_{th}$ and $0\leq t < t_0 $) in terms of the inelastic cross section 
$\sigma_{inel}(s)$. Following \cite{Martin2010} we prove that the bound is a convex function of the inelastic cross section. 
The absorptive part  has the partial wave expansion,
\begin{equation}
  F_s(s,t)\equiv A(s,t)=\frac{ \sqrt s }{4k} \sum_{l=0}^{\infty} (2l+1)P_l (z)Im a_l (s).
\end{equation}
 
The corresponding expansion of the inelastic cross section is,
\begin{equation}
 \sigma _{inel} (s) 
 =\frac{4 \pi}{k^2} \sum _{l=0}^{\infty} (2l+1)\big ( Im a_l(s)-| a_l (s) |^2 \big )\>.
\end{equation}

Actually we shall vary $Im a_l(s)$ subject to the positivity restrictions (due to unitarity), 
\begin{equation}
 Im a_l(s) \geq 0
\end{equation}
 
to minimise $ A(s,t) $, given ,
\begin{equation}
 \sigma _{inel,im} (s) \equiv
 \frac{4 \pi}{k^2} \sum _{l=0}^{\infty} (2l+1)\big ( Im a_l(s)-(Im a_l (s) )^2 \big )\>.
\end{equation}
The bound will be seen to be an increasing function of $\sigma _{inel,im} (s)$. Further,
\begin{equation}
 \sigma _{inel,im} (s) \geq \sigma _{inel} (s);
\end{equation}
therefore the bound will still hold when we replace $ \sigma _{inel,im} (s)$ by the experimentally 
accessible $\sigma _{inel} (s) $. We work at a fixed $s$; so, unless specially needed, we shall suppress 
the $s$-dependence of $Im a_l (s)$, $\sigma _{inel,im} (s) $ and $ \sigma _{inel} (s) $. 
The Lagrange multiplier method with positivity constraints on partial waves gives the variational solution 
$( Im a_l)_0 $ ,
\begin{eqnarray}
 (Im a_l)_0 =\frac{1}{2} \big (1-\frac{P_l (z) } {P_{\lambda }(z) } \big ),\> l \leq L; L\leq \lambda <L+1 \nonumber \\
( Im a_l)_0 = 0, \> l > L \> ; 
\end{eqnarray}
where the integer $L$ and the non-negative fraction $\lambda -L $, are to be determined so as to reproduce the 
given $\sigma _{inel,im} $ ; here $P_{\lambda }(z) $ for non-integer $\lambda $ and $z\geq 1$ is defined by,
\begin{equation}
 P_{\lambda }(z)=\frac{1}{\pi} \int _0 ^\pi (z+\cos \phi \sqrt{z^2-1} )^\lambda d\phi \> .
\end{equation}
 
If $A_0(s,t)$ denotes the absorptive part with partial waves $( Im a_l)_0 $ and 
$A(s,t)$ that with arbitrary positive partial waves with the given $ \sigma _{inel,im} (s)$ ,we obtain 
by direct subtraction,
\begin{eqnarray}
 4(k/\sqrt{s})\big (A(s,t) -A_0(s,t) \big )\nonumber \\
 = P_\lambda (z) \sum _{l=0}^{L} (2l+1)\big ( Im a_l-(Im a_l )_0 \big )^2  \> +\nonumber \\
 P_\lambda (z)\sum _{l=L+1}^{\infty}(2l+1)\big ( (Im a_l)^2+Im a_l \big (\frac{P_l (z) } {P_{\lambda }(z) }-1 \big ) \big) \nonumber \\
 \geq 0.
\end{eqnarray}
The last inequality follows because for $z\geq 1$, and $\lambda \geq 0 $ ,$ P_\lambda (z)$ is an increasing function of $\lambda $.
We then have,

\begin{eqnarray}
 4(k/\sqrt{s})A(s,t) \geq  \sum_{l=0}^{L} (2l+1) P_l (z) \frac{1}{2} \big (1-\frac{P_l (z) } {P_{\lambda }(z) } \big ) \nonumber \\
 \equiv A(\lambda),
 \end{eqnarray}
where,
\begin{eqnarray}
 \sigma _{inel.im} \frac{ k^2}{4\pi}   = \sum_{ l=0}^{l=L } (2l+1)(1/4)\big (1-\big (\frac{P_l (z) } {P_{\lambda }(z) }\big)^2 \big ) \nonumber \\
 \equiv \Sigma _I(\lambda ) .
\end{eqnarray}

Note that $\Sigma _I(\lambda ) $ and $  A(\lambda)$ are monotonically increasing continuous functions of $\lambda $; hence $\lambda $
and $A(\lambda) $ may be considered functions of $\Sigma _I $, and 
\begin{equation}
 dA/d\Sigma _I =\big ( (dA/d\lambda )/ (d\Sigma _I / d\lambda   ) \big )=  P_{\lambda }(z) \>,
 \end{equation}
which is always positive, and also continuous at integer $\lambda$ although $(dA/d\lambda )$ and $(d\Sigma _I / d\lambda   )  $
are discontinuous there. Hence,
\begin{equation}
 d^2 A/d\Sigma _I ^2 = \big ( (d  P_{\lambda }(z) /d\lambda )/ (d\Sigma _I / d\lambda ) \big ) \> > 0, 
\end{equation}
which is discontinuous at integer $\lambda $, but always positive. This completes the proof that $A(\lambda (\Sigma _I ) ) $ 
is a convex function of $\Sigma _I $; i.e. at a given $s$ the lower bound on $A(s,t)$ is an increasing and convex fuction of 
$\sigma _{inel.im} $, and hence of $\sigma _{inel} $.

\noindent {\bf III. Explicit evaluation of the bound}. 
Explicitly, 
\begin{equation}
 A(\lambda (\Sigma _I ) ) =\int _0 ^{\Sigma _I } \frac{dA}{d\Sigma _I '} d\Sigma _I ' = \int _0 ^{\Sigma _I } P_{\lambda '}(z) d\Sigma _I ',
 \end{equation}
where $\lambda ',L' $ corresponds to the value $\Sigma _I ' $ of $\sigma _{inel.im} k^2/(4\pi)$.  When $0<\Sigma _I '< (1-z^{-2})/4 \equiv 
\Sigma _I (1) $ we get $L'=0$ and the corresponding part of the integral can be evaluated exactly. Hence, 
\begin{equation}
 A(\lambda (\Sigma _I ) ) =  (1-z^{-1})/2 + \int _{ \Sigma _I (1)} ^{\Sigma _I } P_{\lambda '}(z) d\Sigma _I ' .
\end{equation}
In the remaining integral $L'\geq 1$ and we shall prove that for $\Sigma _I ' \geq \Sigma _I (1) $, 
\begin{equation}
 (\lambda ') ^2 \geq 4 \Sigma _I '.
\end{equation}

Proof. From the partial wave expansion for $ \Sigma _I(\lambda ')\equiv \Sigma _I ' $, we obtain 
 \begin{equation}
4 \Sigma _I '\leq (L')^2+(2L'+1)\big (1-\big (\frac{P_{L'} (z) } {P_{\lambda '}(z) }\big)^2 \big ).
\end{equation}
The integral representation for $P_{\lambda }(z)$ given before yields, for $z > 1$,
\begin{eqnarray}
 \big (\frac{P_{L'} (z) } {P_{\lambda '}(z) }\big)^2 &\geq& \exp \big ( -2( \lambda ' -L' ) \ln (z+\sqrt{z^2-1 } \>)  \big ) \nonumber \\
  &\geq& 1-2( \lambda ' -L' ) \ln (z+\sqrt{z^2-1 }\> ),
\end{eqnarray}
where the last inequality uses $ \exp (-x) \geq 1-x$. At high energies $\ln (z+\sqrt{z^2-1 } \>) $ goes to zero, and we 
assume moderately high energies ($k>6 m_{\pi} $ ) such that, with $t< 4 m_{\pi}^2 $,
\begin{equation}
 \ln (z+\sqrt{z^2-1 } \>) < 1/3 .
\end{equation}
Then,
\begin{eqnarray}
 4 \Sigma _I '&\leq & (L')^2+(2L'+1)2( \lambda ' -L' ) \ln (z+\sqrt{z^2-1 }\> ) \nonumber \\
 &\leq &  (\lambda ') ^2 - ( \lambda ' -L' )\big( \lambda ' +L' -(2/3) ((2L'+1) )   \big) \nonumber \\
 &\leq & (\lambda ') ^2 \>, for \> L' \geq 1 ,
\end{eqnarray}
which completes the proof.

Since $ P_{\lambda '}(z)$ is an increasing function of  $\lambda '$, we obtain,
\begin{equation}
 A(\lambda (\Sigma _I ) ) =  (1-z^{-1})/2 + \int _{ \Sigma _I (1)} ^{\Sigma _I } P_{\sqrt{4 \Sigma_I ' }}(z) d\Sigma _I ' .
\end{equation} 
Using the integral representation for $ P_{\lambda }(z) $ given before and the analogous representation 
\begin{equation}
 I_0(z)=\frac{1}{\pi} \int _0 ^\pi \exp (z\cos \phi ) d\phi \> 
\end{equation}
for the modified Bessel function , and the elementary inequality 
$$
 \ln ( (z+\cos \phi \sqrt{z^2-1}) \geq (\cos \phi ) \ln ( (z+ \sqrt{z^2-1}),\>z>1
$$
we obtain \cite {Martin2010},
\begin{eqnarray}
  P_{\lambda }(z) \geq I_0 \big (  \lambda  \ln ( (z+ \sqrt{z^2-1}) \big)  \\
 \geq I_0 \big (\lambda \sqrt{z^2-1 } \>/z  \big ), \> z>1.
\end{eqnarray}
Substituting the above inequalities ,the integral over $\Sigma_I ' $ in the expression for the lower bound can be evaluated exactly, 
and we have the exact result (without any high energy approximation),

\begin {eqnarray}
4(k/\sqrt{s})A(s,t) \geq  A (\Sigma _I ) >  \frac{(1-z^{-1})}{2} \nonumber \\
 +(1/2) \big( \ln ( (z+ \sqrt{z^2-1}) )  \big)^{-2} \big[ x' I_1(x') \big] \big|_{x'=u_1 }^{x'=u } ,\\
u\equiv \ln ( (z+ \sqrt{z^2-1}))  \sqrt {k^2 \sigma_{inel }/\pi } ; \nonumber \\
u_1 \equiv \ln ( (z+ \sqrt{z^2-1})) \sqrt {1-z^{-2 }},
\end {eqnarray}
and the slightly weaker but simpler result,
\begin {eqnarray}
  A (\Sigma _I ) &>& \frac{z^2 } {2(z^2-1) } \frac{\sqrt{4 \Sigma_I (z^2-1) }\> } {z } I_1 \big (\frac{\sqrt{4 \Sigma_I (z^2-1) }\> } {z } \big ) 
  \nonumber \\
  &+& \frac{(1-z^{-1})}{2} -\frac{1} {2} I_1 \big ( \frac {z^2-1 } {z^2 }    \big ).
\end {eqnarray}
Note that at high energies $z-1 \rightarrow 0 $, the last two terms give only a small positive contribution,
\begin{equation}
 \frac{(1-z^{-1})}{2} -\frac{1} {2} I_1 \big ( \frac {z^2-1 } {z^2 }    \big ) \approx (z-1)^2 /4 \>, z-1 \rightarrow 0 .
\end{equation}
Hence, at sufficiently high energies, but without any high energy approximation, we have the bound given by Eqs. (37)-(38) and the slightly 
weaker but simpler bound,
\begin{eqnarray}
  A(s,t) &>& \frac {k \sqrt {s} } {4t }   \frac{z^2 } {z+1 } x I_1(x),\> \nonumber \\
  x&\equiv& \sqrt { \frac{z+1 } {2z^2 } }  \sqrt { \frac{t \sigma_{inel }(s) } {\pi } } .
\end{eqnarray}

\noindent {\bf IV. Bound on energy averaged inelastic cross section}.

 Multiplying by $s^{-3}$ and integrating over $s$, we obtain a lower bound on $C_{(s_1,s_2)}(t)$ which is the contribution 
 from $s_1$ to $s_2$ to $C(t)$,
 \begin{eqnarray}
  C_{(s_1,s_2)}(t) &\equiv& \int_{s_1}^{s_2} A(s',t)\frac {ds' } {(s')^3 } \nonumber \\
  &\geq& \int_{s_1}^{s_2}\frac {ds' } {(s')^2 } \frac {\sqrt {(1-4/s_1 ) }} {16t }x_1' I_1(x_1'),\> 
 \end{eqnarray}
where $s_2 > s_1$, and we used $2k' /\sqrt {s'}  >\sqrt {(1-4/s_1 ) } $, and $2z_1^2 /(z_1+1 )\geq 2z^2 /(z+1) \geq 1 $ for $s'$ in the 
interval $(s_1,s_2)$, and
\begin{equation}
 z_1 \equiv \frac {s_1-4+2t}{s_1-4}, \> x_1'\equiv \sqrt { \frac{z_1+1 } {2z_1^2 } }  \sqrt { \frac{t \sigma_{inel }(s') } {\pi } } .
\end{equation}
The lower bound on $C{(s_1,s_2)}(t)$ is an average with the normalized weight function
\begin{equation}
 \rho (s')=\frac{s_1 s_2}{(s_2-s_1)(s')^2}
\end{equation}
of an integrand which is a convex function of $\sigma_{inel }(s') $ .The convexity is readily proved; using $(xI_1(x))'= xI_0(x)$, and denoting,
\begin{equation}
 t_1=t \frac{z_1+1 } {2z_1^2 }= t \frac {(s_1-4)(s_1-4+t) } {(s_1-4+2t)^2 }< t,
 \end{equation}
 we have,
\begin{equation}
\frac {d \big(x_1' I_1(x_1')\big )} {d \sigma_{inel }(s') } = \frac{t_1 } {2\pi } I_0 \big(\sqrt { \frac{t_1 \sigma_{inel }(s') } {\pi } } \big).
\end{equation}
Since the right-hand side is an increasing function of $ \sigma_{inel }(s')$ ,we get the convexity property,
\begin{equation}
 \frac {d^2 \big (x_1' I_1(x_1')\big )} {d \sigma_{inel }(s')^2 } > 0.
\end{equation}
Since the average of a convex function is greater than the convex function of the average \cite{Hardy} ,we have the bound
\begin{equation}
 C_{(s_1,s_2)}(t) \geq \frac{s_2-s_1 } {16ts_1s_2 }\sqrt {(1-4/s_1 ) }\> x_1 I_1(x_1)\>,
\end{equation}
where,
\begin{equation}
 x_1\equiv  \sqrt { \frac{t_1 \bar{\sigma}_{inel }(s_1,s_2) } {\pi } },
\end{equation}

 \begin{equation}
  \bar{\sigma}_{inel }(s_1,s_2) \equiv \int_{s_1}^{s_2}ds' \rho (s')\sigma_{inel } (s').
 \end{equation}
To get bounds on $ \bar{\sigma}_{inel }(s,\infty) $ , and  $\bar{\sigma}_{inel }(s, 2s)$, we just choose the corresponding values for 
$(s_1,s_2) $.Thus we obtain, without any asymptotic approximations in s,

\begin{eqnarray}
 & &C_{(s,\infty)}(t) \geq \frac{1 } {16ts }\sqrt {(1-4/s ) }\> x_1 I_1(x_1)\>, \\
& &  x_1\equiv  \sqrt { \frac{t_1 \bar{\sigma}_{inel }(s,\infty) } {\pi } },
\end{eqnarray}
and,
\begin{eqnarray}
& & C_{(s,2s)}(t) \geq \frac{1 } {32ts }\sqrt {(1-4/s ) }\> x_2 I_1(x_2)\>, \\
& &  x_2\equiv  \sqrt { \frac{t_1 \bar{\sigma}_{inel }(s,2s) } {\pi } },
\end{eqnarray}
where 
\begin{equation}
 t_1= t \frac {(s-4)(s-4+t) } {(s-4+2t)^2 }< t.
 \end{equation} 
 
{\bf Asymptotic bounds }
Since we want asymptotic upper bounds on the energy averaged inelastic cross sections, we can assume without loss of 
generality that the arguments $x_1, x_2$ of the modified Bessel functions tend to infinity , and obtain,
\begin{eqnarray}
& & 16st C_{(s,\infty)}(t) \sqrt{2 \pi } > \big (\sqrt{\xi_1 }\exp{\xi_1 }\big )(1 + O(1/\xi_1)), \\
& & \xi_1=  \sqrt { \frac{t \bar{\sigma}_{inel }(s,\infty) } {\pi } },
\end{eqnarray}
and 
\begin{eqnarray}
& & 32st C_{(s,2s)}(t) \sqrt{2 \pi } > \big (\sqrt{\xi _2 }\exp{\xi _2 }\big )(1 + O(1/\xi_2)), \\
& & \xi_2=  \sqrt { \frac{t \bar{\sigma}_{inel }(s,2s) } {\pi } }.
\end{eqnarray}
We now use the elementary lemma proved in \cite{Martin2010},
\cite {Martin-Roy2014} ,

Lemma. If $\xi >1 $, and $ y \geq \sqrt{\xi }\exp{\xi } $, then, 
\begin{equation}
 \xi< f(y)\equiv \ln{ y} -(1/2)\ln {\big (\ln{ y}-\frac{1}{2}\ln {\ln{ y} } \big )}.
\end{equation}
 
 With $f(y)$ as defined above, we obtain the asymptotic bounds, 
\begin{equation}
 \bar{\sigma}_{inel }(s,\infty) \leq _{s\rightarrow \infty }\frac{\pi } {t } \big ( f(4s/s_0) \big )^2\> ,
\end{equation}
where,
\begin{equation}
 \frac{1 } { s_0}=4t C_{(s,\infty)}(t) \sqrt{2\pi },\> t=4m_\pi ^2 -\epsilon , 
\end{equation}

and,
\begin{equation}
 \bar{\sigma}_{inel }(s,2s) \leq _{s\rightarrow \infty }\frac{\pi } {t } \big ( f(8s/s_0) \big )^2\> .
 \end{equation}

 Notice that the coefficients of $ (\ln s)^2 $ in these bounds on the inelastic cross section are one-fourth of those in 
 the corresponding bounds on the total cross section at high energies , and the scale factors in the inelastic case are also one-fourth of 
 those in the corresponding total cross section bounds \cite {Martin-Roy2014},
\begin{eqnarray}
& & \bar{\sigma}_{tot }(s,\infty) \leq _{s\rightarrow \infty }\frac{4\pi } {t } \big ( f(s/s_0) \big )^2\> ,\nonumber\\
& & \bar{\sigma}_{tot }(s,2s) \leq _{s\rightarrow \infty }\frac{4\pi } {t } \big ( f(2s/s_0) \big )^2\> .
\end{eqnarray}

In the case of pion-pion scattering, we may remove the unknown $\epsilon$ in $t=4m_\pi ^2 -\epsilon $ rigorously by using 
absolute bounds on the D-wave below threshold derived in  \cite {Martin-Roy2014}, or use phenological inputs on the D-wave 
scattering length and set $\epsilon =0  $. The main qualitative difference from the 
 case of non-identical particles is that only even  partial waves occur in $\pi^0 \pi^0 $ scattering, We first show that inspite of this difference, 
 the bounds of this section at moderate energies as well as the asymptotic bounds on inelastic cross sections hold for $\pi^0 \pi^0 $ scattering.

\noindent {\bf V. Bounds on pion-pion inelastic cross sections}. We shall exploit iso-spin invariance. 
\begin{eqnarray}
F^{\pi^0 \pi^0\rightarrow \pi^0 \pi^0}= \frac{1}{3} F^0 + \frac{2}{3}F^2 \nonumber \\
=\frac{ \sqrt s }{4k} \sum_{l=0,2,}^{\infty} (2l+1) 2 a_l ^{\pi^0 \pi^0\rightarrow \pi^0 \pi^0}(s)P_l (z), \nonumber \\
F^{ \pi^0 \pi^0 \rightarrow \pi^+ \pi^- }=\frac{1}{3} F^0 -\frac{1}{3}F^2 \nonumber \\
=\frac{ \sqrt s }{4k} \sum_{l=0,2,}^{\infty} (2l+1) \sqrt{2} a_l ^{\pi^0 \pi^0\rightarrow \pi^+ \pi^-}(s)P_l (z).
\end{eqnarray}
Unitarity then implies,
\begin{eqnarray}
 Im a_l ^{\pi^0 \pi^0\rightarrow \pi^0 \pi^0} &\geq & | a_l ^{\pi^0 \pi^0\rightarrow \pi^0 \pi^0} |^2  +  | a_l ^{\pi^0 \pi^0\rightarrow \pi^+ \pi^-} |^2 \nonumber \\
& = &\frac{1}{3}| a_l^0 (s) |^2 +\frac{2}{3}| a_l^2 (s) |^2 .
\end{eqnarray}
Hence we define the inelastic cross section considering $ \pi^0 \pi^0\rightarrow \pi^+ \pi^- $ also as an elastic channel,
\begin{eqnarray}
 \sigma_{inel }^{ \pi^0 \pi^0 }\equiv  \sigma_{tot }^{ \pi^0 \pi^0} -  \sigma^{ \pi^0 \pi^0 \rightarrow \pi^0 \pi^0 } 
 -  \sigma^{ \pi^0 \pi^0 \rightarrow \pi^+ \pi^- } = \frac{8 \pi}{k^2} \times \nonumber \\  
 \sum _{l=0,2,}^{\infty} (2l+1)\big (  \frac{1}{3}(Im a_l^0-| a_l^0 |^2) +\frac{2}{3}(Im a_l^2 -| a_l^2 |^2) \big ) \nonumber.
\end{eqnarray}
Note that, 
\begin{eqnarray}
 \sigma_{inel }^{ \pi^0 \pi^0 } \leq \sigma_{inel,im }^{ \pi^0 \pi^0 }\equiv  \frac{8 \pi}{k^2} \sum _{l=0,2,}^{\infty} (2l+1) \times \nonumber \\  
\big (  \frac{1}{3}(Im a_l^0-(Im a_l^0)^2) +\frac{2}{3}(Im a_l^2 -(Im a_l^2)^2) \big ).
\end{eqnarray}

As before, we vary the $ Im a_l^I $ subject to positivity constraints , and the given $\sigma_{inel,im }^{ \pi^0 \pi^0 }$ to minimise the 
aborptive part, 
\begin{eqnarray}
  A^{ \pi^0 \pi^0 \rightarrow \pi^0 \pi^0 }(s,t)=\frac{ \sqrt s }{4k} \sum_{l=0,2}^{\infty} (2l+1)\times \nonumber \\
  P_l (z)2\big (  \frac{1}{3}(Im a_l^0) +\frac{2}{3}(Im a_l^2 ) \big ).
\end{eqnarray}
The minimum is reached when 
\begin{eqnarray}
 & &Im a_l^0 =Im a_l^2=\frac{1}{2} \big (1-\frac{P_l (z) } {P_{\lambda }(z) } \big ),\> l \leq L; \nonumber \\
 & &Im a_l^0 =Im a_l^2= 0, \> l > L \> ,L\leq \lambda <L+2 . 
\end{eqnarray}
The minimum is an increasing and a convex function of  $\sigma_{inel,im }^{ \pi^0 \pi^0 }$. The lower bound on the absorptive part therefore 
remains valid if we replace   $\sigma_{inel,im }^{ \pi^0 \pi^0 }$ by $\sigma_{inel }^{ \pi^0 \pi^0 }$. Again, defining
\begin{eqnarray}
 4(k/\sqrt{s})A^ { \pi^0 \pi^0 \rightarrow \pi^0 \pi^0 }(s,t) \equiv A^{ \pi^0 \pi^0 }(\lambda), \\
  \sigma _{inel}^{ \pi^0 \pi^0 } \frac{ k^2}{4\pi}   \equiv \Sigma _I^{ \pi^0 \pi^0 }(\lambda ) ,
\end{eqnarray}
we prove that if $ln (z+\sqrt{z^2-1 } ) <1/6$, which holds at moderately high energies, $\Sigma _I^{ \pi^0 \pi^0  }(\lambda )< \lambda^2/4 $ if $L\geq 2$. Finally,
proceeding as for non-identical particles, we obtain a bound without any high energy approximations, 
\begin{eqnarray}
 A^{ \pi^0 \pi^0 }(\lambda (\Sigma _I^{ \pi^0 \pi^0 } ) ) \geq  (1-1/P_2(z)) + \nonumber \\
 \int _{ \Sigma _I^{ \pi^0 \pi^0 } (2)} ^{\Sigma _I^{ \pi^0 \pi^0 } } P_{\sqrt{4 \Sigma_I ' }}(z) d\Sigma _I ' .
\end{eqnarray}
This yields the exact bound,
\begin {eqnarray}
&&4(k/\sqrt{s})A^ { \pi^0 \pi^0 }(s,t)\geq    (1-1/P_2(z)) +\nonumber \\
&& (1/2)\alpha (z)^{-2} \big[ x' I_1(x') \big] \big|_{x'=v_1 }^{x'=v } ,\\
&&\alpha (z)\equiv \ln ( (z+ \sqrt{z^2-1})), v\equiv \alpha(z)  \sqrt {\frac{k^2}{\pi} \sigma_{inel }^{ \pi^0 \pi^0 } } ; \nonumber \\
&&v_1 \equiv \alpha(z)  \sqrt   {2(1-P_2(z)^{-2 })};
\end {eqnarray}
a slightly weaker bound is obtained by replacing $\alpha(z)$ by the smaller quantity $ \sqrt{z^2-1 }/z $, and noting that 
$ (1-1/P_2(z))- (1/2)\alpha (z)^{-2} \big[ x' I_1(x') \big] \big|_{x'=v_1 }$ is then positive at moderately high energies.
Thus we obtain a slightly weaker but rigorous bound,
  \begin{eqnarray}
  A^{ \pi^0 \pi^0 }(s,t) &>& \frac {k \sqrt {s} } {4t }   \frac{z^2 } {z+1 } x I_1(x),\> \nonumber \\
  x&\equiv& \sqrt { \frac{z+1 } {2z^2 } }  \sqrt { \frac{t \sigma_{inel }^{ \pi^0 \pi^0 }(s) } {\pi } } 
\end{eqnarray}
which is identical to the result given earlier for non-identical particles. Therefore the asymptotic bounds of the last section 
on energy averages of inelastic cross sections also hold for $\pi^0 \pi^0  $ scattering.

\noindent {\bf VI. Absolute bounds on  $\pi^0 \pi^0$  inelastic cross sections}.In \cite{Martin-Roy2014} we derived absolute bounds 
on $  \pi^0 \pi^0$ D-waves below threshold and on $C (t)$ in terms of pion-mass alone, for $0<t<4$. In particular,

\begin{equation}
f_2(t)<_{t\rightarrow 4- } \frac{4-t}{120}\big (34+6.25(4-t)+O(4-t)^2 \big ).
\end{equation}
and
\begin{equation}
  C^{\pi^0 \pi^0\rightarrow \pi^0 \pi^0 }(t)<_{t\rightarrow 4- }\frac{17 \pi}{4(4-t)},
\end{equation}
where $f_l(s)=\frac{ \sqrt s }{4k}  2 a_l ^{\pi^0 \pi^0\rightarrow \pi^0 \pi^0}(s)$. We now use these 
rigorous bounds in conjunction with the asymptotic bounds on inelastic $ \pi^0 \pi^0 $ cross sections to remove 
the unknown $ \epsilon =4-t $ in these bounds.The price to pay for the rigour is that we cannot choose $t=4$. 
For the upper bound on  $\bar{\sigma}_{inel }^{\pi^0 \pi^0}(s,\infty)  $, the optimum choice is,

\begin{equation}
 \epsilon=4-t = \frac{8}{\ln (s/s_1)}, 
\end{equation}
which yields , 
 \begin{eqnarray}
\bar{\sigma}_{inel }^{\pi^0 \pi^0} (s,\infty)  \leq _{ s\rightarrow \infty }  (\pi /4) (m_{\pi })^{-2} \times \nonumber \\
\big(\ln (s/s_1)+(1/2)\ln \ln (s/s_1) +1 \big)^2   +O(\ln \ln (s/s_1) ) .   
\end{eqnarray}
 where the scale factor $s_1 $ is given by
\begin{equation}
 1/s_1= 34\pi \sqrt{2\pi }\>m_{\pi }^{-2}, 
\end{equation}
and is one-fourth of that for the total cross section case\cite{Martin-Roy2014}.  
 For the upper bound on  $\bar{\sigma}_{inel }^{\pi^0 \pi^0}(s,2s)  $, 
 the optimum choice is,

\begin{equation}
 \epsilon=4-t = \frac{8}{\ln (s/s_2)}, 
\end{equation}
which yields , 
 \begin{eqnarray}
\bar{\sigma}_{inel }^{\pi^0 \pi^0} (s,2s)  \leq _{ s\rightarrow \infty } (\pi /4) (m_{\pi })^{-2} \times \nonumber \\
\big(\ln (s/s_2)+(1/2)\ln \ln (s/s_2) +1\big)^2  + O(\ln \ln (s/s_2) )  .   
\end{eqnarray}
 where the scale factor $s_2 $ is given by
\begin{equation}
 1/s_2= 2 /s_1= 68 \pi \sqrt{2\pi }\>m_{\pi }^{-2}. 
\end{equation}
 These are bounds from first principles on a cross section fundamental in strong interaction physics. But for phenomenological comparisons 
 it is more useful to use some crossed channel low energy data to get stronger bounds , particularly on the scale of the logarthm.
 
 {\bf VII. Phenomenological comparisons for Pion-Pion scattering .}
 
 (i) First, the basic lower bound (from unitarity alone) on the absorptive part $A(s,t)$ in terms of the inelastic cross section, given by Eq. (41), 
 or in terms of the total cross section (Eq. (21) of \cite{Martin-Roy2014} ) can be compared directly with phenomenological estimates of the 
 absorptive part at energies where such estimates are available \cite{Leutwyler},
 \cite{Colangelo2015}.
 This can be done for the amplitudes  
 $$F^{\pi^+ \pi^0\rightarrow \pi^+ \pi^0 (s,t)}=
1/2( F^1 + F^2) (s,t),$$
$$F^{\pi^0 \pi^0\rightarrow \pi^0 \pi^0} (s,t)= \frac{1}{3} F^0 + \frac{2}{3}F^2 (s,t), $$
which have positive absorptive parts for $s\geq 4$, $0 < t < 4$. A violation of the bounds will indicate that 
the input absorptive part violates unitarity. 

(ii) Secondly, bounds on energy averages of cross sections in the intervals $(s.\infty)$ and $(s.2s)$ 
in terms of phenomenological inputs for c.m. energies less than $\sqrt{s}$ follow from unitarity , analyticity and crossing .
The crossing relations, 
\begin{eqnarray}
\frac{1}{2}( F^1 + F^2)(s,t)=\frac{1}{3}(F^0-F^2)(t,s), \nonumber \\ 
F^{\pi^0 \pi^0\rightarrow \pi^0 \pi^0} (s,t) =F^{\pi^0 \pi^0\rightarrow \pi^0 \pi^0} (t,s)
\end{eqnarray}
and the Froissart Gribov formula yield, 
\begin{eqnarray}
 C^{\pi^+ \pi^0\rightarrow \pi^+ \pi^0 }_{(s.\infty)}(t=4)= \nonumber \\
 \frac{5\pi } {16 }m_\pi (a_2^0 -a_2^2)\>-C^{\pi^+ \pi^0\rightarrow \pi^+ \pi^0 }_{(4,s)}(t=4),
\end{eqnarray}
and 
\begin{eqnarray}
 C^{\pi^0 \pi^0\rightarrow \pi^0 \pi^0 }_{(s,\infty)}(t=4)= \nonumber \\ 
 \frac{5\pi } {16 }m_\pi (a_2^0 +2a_2^2)\> -C^{\pi^0 \pi^0\rightarrow \pi^0 \pi^0 }_{(4,s)}(t=4).
\end{eqnarray}
Here , as in \cite{Martin-Roy2014}, we defined the $l$-wave scattering lengths $a_l^I$ as the $q \rightarrow 0$ limits of 
the phase shifts $\delta _l ^I (q) $ divided by $q^{2l+1}$ where $q$ is the c.m. momentum . The Bern group \cite{Colangelo2000} 
already has estimates of the $D$-wave scattering lengths, and has recently obtained  \cite{Colangelo2015a} estimates 
of the absorptive part integrals upto $\sqrt{s}=1,6 GeV$ ,
$$C^{\pi^+ \pi^0\rightarrow \pi^+ \pi^0 }_{(4m_\pi^2,s)}(t=4) = 1.48 \times 10^{-3} ,$$  
$$C^{\pi^0 \pi^0\rightarrow \pi^0 \pi^0 }_{(4m_\pi^2,s)}(t=4)= 2.031 \times 10^{-3}. $$
Hence the bounds on energy averaged cross sections 
$\bar{\sigma}_{inel }^{\pi^0 \pi^0} (s,2s) $ and $\bar{\sigma}_{inel }^{\pi^+ \pi^0} (s,2s) $, as well as  
$\bar{\sigma}_{inel }^{\pi^0 \pi^0} (s,\infty) $ and $\bar{\sigma}_{inel }^{\pi^+ \pi^0} (s,\infty)$  implied by 
Eqs. (51)-(55) can be directly tested against the corresponding experimental values.

(iii) Thirdly, explicit asymptotic bounds on the averages of the inelastic cross section in the intervals $(s,\infty) $ and $(s,2s)$ 
are given  by Eqs. (61) and (63) , and in terms of the corresponding averages of the total cross section , 
in terms of a scale parameter $s_0$ ; $s_0$ is given by Eq. (62) in terms of 
$C_{(s,\infty)}$, an integral over absorptive parts in the interval $(s,\infty)$ . Substituting the values of the D-wave scattering lengths 
given by \cite{Colangelo2000},
\begin{equation}
 a_2^0\approx 0.00175 m_\pi ^{-5 }\>;a_2^2\approx 0.00017 m_\pi ^{-5 },
\end{equation}
we have, choosing for $s$ a value up to which absorptive parts can be reliably estimated,
\begin{eqnarray}
 \pi^0 \pi^0: \>s_0^{-1}= m_\pi ^{-2} 16 \sqrt{2\pi}\nonumber\\
 \times \big (2.05\times 10^{-3}-C^{\pi^0 \pi^0\rightarrow \pi^0 \pi^0 }_{(4,s)}(t=4) \big),
\end{eqnarray}
\begin{eqnarray}
\pi^+ \pi^0 :\>s_0^{-1}= m_\pi ^{-2} 16 \sqrt{2\pi}\nonumber\\
 \times \big (1.55\times 10^{-3}-C^{\pi^+ \pi^0\rightarrow \pi^+ \pi^0 }_{(4,s)}(t=4) \big).
\end{eqnarray}
 These equations give much stronger bounds than the absolute bounds. E.g. using only positivity of 
 $C^{\pi^0 \pi^0\rightarrow \pi^0 \pi^0 }_{(4,s)}(t=4) $ we get, 
 $$ s_0 \geq 12 m_{\pi}^2,$$
which is $800$ times the absolute bound $ s_0 \geq .015 m_{\pi}^2$ . As the absorptive part integrals in the $D$-wave 
scattering length sum rules are rapidly convergent, even for the moderate value of $\sqrt{s}=1,6 GeV$  , Colangelo et al \cite{Colangelo2015a} 
obtained a further big  improvement in the values of the scale factor, when 
phenomenological values of absorptive parts upto $\sqrt{s}=1,6 GeV$ are utilised , 
\begin{eqnarray}
\pi^0 \pi^0 :\>s_0 \geq 1312 m_\pi ^2 \nonumber\\
\pi^+ \pi^0 :\>s_0 \geq 356 m_\pi ^2 ,
 \end{eqnarray}
which are not very far from the scale factors used in phenomenological fits \cite{Colangelo2015}.
We should remember that the phenomenological values may be dependent on the particular parametrisation 
used to fit experimental cross sections. The implicit bounds (51)-(55)  discussed in (i) and (ii) are without asymptotic approximations,
and therefore can be compared directly with experiment.

 {\bf VIII. Concluding Remarks.}
 
In this paper on inelastic cross sections and the previous one 
on total cross sections \cite {Martin-Roy2014} we believe to have put the Froissart
Bound on a solid ground, by using the notion of average cross sections
which avoids completely the problem of the scale in the Froissart bound.
These averages can be chosen rather arbitrarily  but once you have
chosen one you must stick to it. The simplest averages that we use are
the ones from $s$ to infinity and from $s$ to $2s$. The averaging interval must
be sufficiently large if one wants to preserve the coefficients
appearing in the Lukaszuk-Martin bound. The only unknown is the value of
a certain integral on the absoptive part for some positive $t$. In the
special case of pion-pion scattering all unknown constants are
eliminated. The advantage of introducing the bound on the inelastic
cross section is that, asymptotically, it is 4 times smaller than the one on the total
cross section. So if you accept to believe that the elastic
cross-section cannot be larger than the inelastic cross section, the
limiting case being an expanding black disk you gain a factor 2 on the
bound on the total cross section. However, not everybody agrees with
this, for instance Troshin and Tyurin \cite{Troshin-Tyurin} believe that at high
enrergy the scattering amplitude is dominantly elastic. It is tempting
to make a rather daring and non-rigorous suggestion: if the amplitude
is essentially elastic  ( a small inelastic part is unavoidable
according to well known theorems) then the effective large Lehmann
ellipse has a right extremity at $t= 16 m_{\pi}^2 $, and the Froissart
bound is divided by 4.

    Anyway a factor of 2 or 4 is not sufficient to bring the absolute bounds 
near the experimental values \cite{ISR},\cite{SppbarS},\cite{Tevatron} including the most recent experiments at LHC \cite{LHC}
, which indicate a definite increase of the cross-sections compatible with a $(\ln(s))^2$ behaviour. 
There is little doubt that this trend will continue when LHC reaches higher energies.Towards quantitative improvement we may 
find unitarity bounds on the energy averages of the inelastic cross section given the total cross section 
as an input , in addition to absorptive part integrals at positive $t$ \cite{Martin-Roy2015}.

However , as explained in Sec. VII above,if we are prepared to make phenomenological inputs such as the $D$-wave scattering lengths 
and low energy absorptive parts , the situation with respect to experimental comparisons 
 improves dramatically \cite{Colangelo2015}.

What can we do on the theoretical side? In the case of pion pion
scattering, Kupsch\cite{Kupsch} has constructed an amplitude , crossing-symmetric
satisfying "inelastic" unitarity and saturating the Froissart bound 
\cite{Kupsch} ,but
he does not give numbers. The result of Gribov \cite{Gribov} shows the
importance of satisfying elastic unitarity in the "elastic strips" ,
\cite{Martin-Richard}. This might help, but we dont know how, and
there is the problem of finding people interested in working on this.

 We acknowledge very stimulating correspondence with  I. Caprini,  G. Colangelo, J. Gasser and H. Leutwyler of 
 the University of Bern , and thank them for sending us their unpublished results, Eq. (90),  on phenomenological 
 evaluation of the bounds on scale factors for pion-pion scattering. We also thank 
 Maurice Haguenauer for giving us up to date references on the work of the Atlas Alpha collaboration. 
 SMR would like to thank the Indian National Science Academy for an INSA senior scientist grant.


\begin{thebibliography}{99}
\bibitem{Martin-Roy2014} A. Martin and S. M. Roy , Phys. Rev. {\bf D89}, 045015 (2014).
\bibitem{Froissart1961} M. Froissart, Phys. Rev. {\bf 123}, 1053 (1961).
\bibitem{Martin1966} A. Martin, Nuov. Cimen. {\bf 42}, 930 (1966).
\bibitem{Wightman} See e.g. R. F. Streater and A. S. Wightman, ``PCT. Spin and Statistics and all that'' 
(2nd revised printing , Benjamin, New York (1978)) reprinted by Princeton University Press (2000).
\bibitem{Epstein} H. Epstein,V. Glaser and A. Martin, Comm. Math. Phys., {\bf 13}, 257 (1969). 
\bibitem{Haag} R. Haag and D. Kastler, J. Math. Phys., {\bf 5} 848 (1964); D. Ruelle, Helv. Phys. Acta 
{\bf 35}, 147 (1962);H. Araki, ``Mathematical Theory of Quantum Fields'', Oxford Univ. Press, Oxford (2000).
\bibitem{Azimov} Y. I. Azimov, Phys. Rev. {\bf D84},056012 (2011).
\bibitem{Zimmermann}
W. Zimmermann, Nuov. Cimen. {\bf 10}, 597 (1958), {\bf 21},249(1961) and {\bf 21},268(1961).
\bibitem{Lehmann1958} H. Lehmann, Nuovo Cimen. {\bf 10}, 579 (1958); Fortschr. Physik {\bf 6} 159 (1959).
\bibitem{Bessis-Glaser1967} J. D. Bessis and V. Glaser, Nuov. Cimen. {\bf 50}, 568 (1967);
G. Sommer, Nuov. Cimen. {\bf A48}, 92 (1967); for reviews, 
see G. Sommer, Fortschritte der Physik, {\bf 18},577 (1970);  A. Martin, CERN-TH/99-110 (1999).  
\bibitem{Jin-Martin1964} Y. S. Jin and A. Martin, Phys. Rev. {\bf 135B}, 1375(1964).
\bibitem{Lukaszuk-Martin1967} L. Lukaszuk and A. Martin, Nuov. Cimen. {\bf 52A}, 122 (1967).
\bibitem{Singh-Roy1970} V. Singh and S. M. Roy, Ann. Phys. {\bf 57}, 461 (1970).
\bibitem{Roy1972} S. M. Roy, Phys. Reports, {\bf 5C}, 125 (1972).
\bibitem{Cheng-Wu1970} H. Cheng and T. T. Wu, Phys. Rev. Letters {\bf 24},1456 (1970);
C. Bourrely, J. Soffer, and T. T. Wu, Phys. Rev. {\bf D19}, 3249 (1979) and Nucl. Phys. 
{\bf B247}, 15 (1984), M. M. Block, Phys. Reports, {\bf 436}, 71 (2006) and references therein.
\bibitem{Martin2009} A. Martin, Phys. Rev. {\bf D80}, 065013 (2009).
\bibitem{WMRS} T. T. Wu, A. Martin, S. M. Roy and V. Singh, Phys. Rev. {\bf D84}, 025012 (2011).
\bibitem{Yndurain} F J. Yndurain, Phys. Letters {\bf 31B},368 (1970); 
A. K. Common, Nuovo Cim. {\bf 69A },115(1970).
\bibitem{Colangelo2000} G. Colangelo, J. Gasser, and H. Leutwyler, Nuclear Physics {\bf B603}, 125(2001); 
B. Ananthanarayan,G. Colangelo, J. Gasser, and H. Leutwyler, Physics Reports {\bf 353}, 207 (2001).
\bibitem{Martin2010} A. Martin, Pomeranchuk prize lecture, ITEP, Moscow (2010), https://cds.cern.ch/record/1591290. 
\bibitem{Hardy} G. H. Hardy,J. E. Littlewood and G. P$\acute{o}$lya, {\it Inequalities},p.74, Cambridge 
University Press (1952).
\bibitem{Martin1967} A. Martin,Nuovo Cimento {\bf 47},265 (1967).
\bibitem{Martin1965} A. Martin, {\it High Energy Physics and Elementary Particles},p.155, IAEA Vienna (1965).
\bibitem{Auberson et al} G. Auberson, L. Epele, G. Mahoux and F. R. A. Simao, Nucl. Phys. {\bf B73 },314 (1974);
Nucl. Phys. {\bf B94 },311 (1975).
\bibitem{Lopez-Mennessier} C. Lopez and G. Mennessier, Phys. Lett. {\bf B58},437(1975), and 
Nucl. Phys. {\bf B 118}, 426(1977).
\bibitem{Leutwyler} I. Caprini, G. Colangelo, J. Gasser and H. Leutwyler, 
Phys.Rev. {\bf D68},  074006 (2003); e-Print: hep-ph/0306122.
\bibitem{Colangelo2015} I. Caprini,G. Colangelo, and H. Leutwyler, Eur.Phys.J. {\bf C72},1860,(2012) and 
arXiv:1111.7160v2 [hep-ph] 7 Aug 2012; I. Caprini,G. Colangelo, J. Gasser and H. Leutwyler, (private communication). 
\bibitem{Colangelo2015a} I. Caprini,G. Colangelo, and H. Leutwyler, private communication (to be published).
\bibitem{Troshin-Tyurin} S. M. Troshin and N. E. Tyurin, Phys. Lett. {\bf 208 B}, 517 (1988).
\bibitem{ISR} U. Amaldi et al, Phys. Lett. {\bf B44},112(1973);S. R. Amendolia et al,Phys. Lett.{\bf B44},119(1973).
\bibitem{SppbarS} M. Bozzo et al, Phys. Lett. {\bf B147},392(1984); C. Augier et al, Phys. Lett. {\bf B316},448(1993).
\bibitem{Tevatron} N. A. Amos et al, Phys. Rv. Lett. {\bf 63},2784 (1989),Phys. Lett. {\bf B243},158(1990), and 
Phys. Rev. Lett. {\bf 68 },2433 (1992).
\bibitem{LHC} Totem Collaboration, EPL {\bf 101 },21004 (2013); Atlas-Alpha collaboration, Nucl. Phys. {\bf B889}, 486-548(2014)
\bibitem{Martin-Roy2015} A. Martin and S. M. Roy, manuscript in preparation.
\bibitem{Kupsch} J. Kupsch, Nuovo Cim. {\bf 71A},85 (1982).
\bibitem{Gribov} V. N. Gribov, {\it Proc. International Conf. on High Energy Physics at Rochester}, 
Edited by E. C. G. Sudershan, J. H. Tinlot and A. C. Melissos (Univ. of Rochester, Rochester, NY, 1960),p. 340.
\bibitem{Martin-Richard} A. Martin and J.-M. Richard, {\it Proc. 'Forward Physics and Luminosity Determination at LHC',
Helsinki 2000 }, K. Haiti et al Eds., p.27 (World Scientific 2001).


\end{thebibliography}
\end{document}